\documentclass[aip,jcp,11pt,reprint,floatfix,superscriptaddress,urlname]{revtex4-1}

\usepackage[utf8]{inputenc}

\usepackage{amsmath}
\usepackage{color}
\usepackage{graphicx}
\usepackage{comment}
\usepackage{caption}
\usepackage{subcaption}
\usepackage[normalem]{ulem}
\usepackage{algorithm}
\usepackage{algpseudocode}
\usepackage{rotating}
\usepackage{amsmath,amssymb}
\usepackage{graphicx}
\usepackage{booktabs}
\usepackage{mathtools}

\usepackage{xcolor}

\newcommand{\bra}[1]{{\big< #1 \big|}}
\newcommand{\ket}[1]{{\big| #1 \big>}}

\DeclareMathAlphabet      {\mathbfit}{OML}{cmm}{b}{it}

\begin{document}

\title{AB-G$_0$W$_0$: A practical G$_0$W$_0$ method without frequency integration based on an auxiliary boson expansion}

\author{Johannes T\"olle}
\email{jtolle@caltech.edu}
\author{Garnet Kin-Lic Chan}
\email{gkc1000@gmail.com}
\noaffiliation
\affiliation{Division of Chemistry and Chemical Engineering, California Institute of Technology, Pasadena, California 91125, USA}

\begin{abstract}
Common G$_0$W$_0$ implementations rely on numerical or analytical frequency integration to determine the G$_0$W$_0$ self-energy, which results in a variety of practical complications. 
Recently, we demonstrated an exact connection between the G$_0$W$_0$ approximation and equation-of-motion (EOM) quantum chemistry approaches [\textit{J. Chem. Phys}, \textbf{158}, 124123 (2023)].
Based on this connection, we propose a new  method to determine G$_0$W$_0$ quasiparticle energies which completely avoids frequency integration and its associated problems.  
To achieve this, we make use of an auxiliary boson (AB) expansion. We name the new approach AB-G$_0$W$_0$ and demonstrate its practical applicability in a range of molecular problems.
\end{abstract}
\maketitle 

\clearpage
The G$_0$W$_0$ method has become a widely used tool for the determination of quasiparticle energies due to its accuracy-to-cost ratio, c.f.~Refs.~\citenum{golze2019gw,bruneval2021gw,marie2023gw}.
The G$_0$W$_0$ self-energy, $\Sigma^\mathrm{G_0W_0}$ can be defined as the following integral
\begin{align}
    \Sigma^\mathrm{G_0W_0} (\mathbf{r},\mathbf{r}',\omega) &= \frac{\mathrm{i}}{2 \pi} \int d\omega'  e^{\mathrm{i}\omega' \eta} G_0(\mathbf{r},\mathbf{r'},\omega + \omega') \nonumber \\
    &\times W_0(\mathbf{r},\mathbf{r}',\omega'),
    \label{eq:freqInt}
\end{align}
with $G_0$ the non-interacting Green's function, $W_0$ the screened-Coulomb interaction, with the above integration along the real axis (with positive infinitesimal $\eta$).
This frequency integral 
is inconvenient due to the poles of $G_0$ and $W_0$ along the real axis \cite{golze2019gw}.
Practical realizations of the G$_0$W$_0$ method, therefore, make use of different techniques to compute this integral.
In ``fully-analytic'' approaches, an analytical expression for Eq.~(\ref{eq:freqInt}) is used in terms of the poles and residues \cite{bruneval2012ionization,van2013gw}.
Alternatively, Eq.~(\ref{eq:freqInt}) is computed by numerical integration. 
In the common analytic continuation (AC) approach, the self-energy is evaluated for imaginary frequencies using an imaginary frequency integral, and then analytically continued to the real-frequency axis, c.f.~Refs.~\citenum{rojas1995space,ren2012resolution,wilhelm2016gw,wilhelm2018toward,zhu2021all}. Another common strategy uses a contour deformation in the complex plane (CD), i.e.~Refs.~\citenum{godby1988self,golze2018core,holzer2019ionized,zhu2021all}, to evaluate the real-frequency integral. 

These different realizations of the frequency integration exhibit different computational scalings and other advantages and disadvantages.
For example, the ``fully-analytic'' approach diagonalizes the random-phase approximation (RPA) eigenvalue problem, leading to an expensive $\mathcal{O}(N_o^3 N_v^3)$ scaling ($N_o$, $N_v$ being the number of occupied and virtual orbitals). In numerical integration approaches, it is common to introduce an auxiliary density fitting basis of dimension $N_\mathrm{aux}$. Then the AC approach scales as $\mathcal{O}(N_o N_v N_\mathrm{aux}^2)$ (which can be further reduced to cubic in system size 
by constructing intermediates in the imaginary time domain \cite{rojas1995space,liu2016cubic,foerster2020low,forster2021gw100,duchemin2021cubic}).
However its applicability is limited to valence states due to the instability of the analytic continuation \cite{vidberg1977solving,golze2018core,duchemin2020robust}.
The CD approach also scales as $\mathcal{O}(N_o N_v N_\mathrm{aux}^2)$ for valence quasiparticle energies, but includes a contribution from the summation over poles along the real frequency axis. For core- or high lying virtual quasiparticle energies, the number of such poles (from the  occupied valence states/virtual valence states) scales as $N_\mathrm{val}$, changing the overall scaling to $\mathcal{O}(N_o N_v N_\mathrm{val} N_\mathrm{aux}^2)$ \cite{golze2018core}.

Because of the above complicating numerical considerations, there has been much recent activity to approximate the G$_0$W$_0$ method without reference to frequency-dependent quantities~\cite{bintrim2021full,quintero2022connections,tolle2023exact,scott2023moment}. 
We have shown recently that an exact frequency-independent formulation of G$_0$W$_0$ is possible in terms of the equation-of-motion formalism of unitary coupled cluster theory~\cite{tolle2023exact}. However, 
the formalism described in Ref.~\citenum{tolle2023exact}, when implemented exactly, has a scaling of $\mathcal{O}(N_o^3 N_v^3)$ and is thus impractical for real applications. In the current work, we describe a simple quartic-scaling implementation of the equation-of-motion formulation of G$_0$W$_0$ that relies on expanding the quasi-bosonic space in an auxiliary bosonic basis that scales linearly with the system size.
We demonstrate numerically that this new G$_0$W$_0$ formulation, termed auxiliary-boson (AB)-G$_0$W$_0$, allows for the practical and unbiased calculation of core and valence quasiparticle energies in molecules of non-trivial size. 

\section{Theory}
\label{sec:Theory}
Within the quasi-boson approximation the products of fermionic particle ($\hat{a}^\dagger_a$/$\hat{a}_a$) and hole ($\hat{a}_i$/$\hat{a}^\dagger_i$) creation/annihilation operators are approximated by bosonic creation ($\hat{b}^\dagger_\nu$) and annihilation ($\hat{b}_\nu$) operators
\begin{align}
    \hat{a}^\dagger_a \hat{a}_i &\approx \hat{b}^\dagger_\nu \\
    \hat{a}^\dagger_i \hat{a}_a &\approx \hat{b}_\nu.
\end{align}
These quasi-bosons satisfy bosonic commutation relations \cite{ring2004nuclear}.
In the following, occupied orbitals are denoted as $i,j,k\dots$, virtual orbitals are denoted as $a,b,c\dots$, general orbital indices are denoted as $p,q,r\dots$, small Latin letters denote fermionic indices, and Greek letters are used for bosonic indices.
As demonstrated in Ref.~\citenum{tolle2023exact}, starting from an electron-boson Hamiltonian
\begin{align}
    \hat{H}^\mathrm{eB} = \hat{H}^\mathrm{e} + \hat{H}^\mathrm{B} + \hat{V}^\mathrm{eB}
    \label{eq:EBHam}
\end{align}
with (real-valued orbitals are assumed throughout)
\begin{align}
    \hat{H}^\mathrm{e} &= \sum_{pq} f_{pq} \{\hat{a}^\dagger_p \hat{a}_q\}\\
    \hat{H}^\mathrm{B} &= \sum_{\nu \mu} A_{\nu \mu} \hat{b}^\dagger_\nu \hat{b}_\mu + \frac{1}{2} \sum_{\nu \mu} B_{\nu \mu} (\hat{b}^\dagger_\nu \hat{b}^\dagger_\mu + \hat{b}_\nu \hat{b}_\mu) \label{eq:HB}\\
    \hat{V}^\mathrm{eB} &= \sum_{pq,\nu} V_{pq \nu} \{ \hat{a}^\dagger_p \hat{a}_q\}  (\hat{b}^\dagger_\nu + \hat{b}_\nu),
    \label{eq:eBHamiltonian}
\end{align}
where $\{ \dots \}$ denotes normal-ordered fermionic operators with respect to the Fermi vacuum,
G$_0$W$_0$ quasiparticle energies
can be obtained using an EOM approach \cite{bintrim2021full}.
In Eq.~(\ref{eq:eBHamiltonian}), $A_{\nu \mu}$ and $B_{\nu \mu}$ denote the direct random-phase-approximation (dRPA) response matrices
\begin{align}
    A_{\nu \mu}  &= A_{ia,jb} = \delta_{ij} \delta_{ab} (\epsilon_a - \epsilon_i) + (ia|bj) 
    \label{eq:AMat}\\
    B_{\nu \mu}  &= B_{ia,jb} = (ia|jb),
    \label{eq:BMat}
\end{align}
and 
\begin{align}
    V_{pq \nu} = V_{pq,ia} = (pq|ia),
\end{align}
with electron repulsion integrals expressed in $(11|22)$ (Mulliken) notation.
To obtain the EOM supermatrix, we first perform a
unitary (canonical) transformation of $\hat{H}^\mathrm{eB}$ to block diagonalize $\hat{H}^\mathrm{B}$, removing the non-boson-number conserving terms,
\begin{align}
    \hat{U}^\dagger \hat{H}^\mathrm{eB} \hat{U} \rightarrow \bar{H}^\mathrm{eB} 
    \label{eq:CT}
\end{align}
with
\begin{align}
    \bar{H}^\mathrm{B} = \sum_{\nu \mu} \bar{A}_{\nu \mu} \hat{b}^\dagger_\nu \hat{b}_\mu + E^\mathrm{c}_\mathrm{dRPA}.
\end{align}
We then build the supermatrices $\mathbf{H}$, $\mathbf{S}$, with matrix elements
\begin{align}
    H_{IJ} &= \langle 0_\mathrm{F}0_\mathrm{B}| [{C}_I, [\bar{H}^\mathrm{EB}, {C}^\dag_J]] |0_\mathrm{F}0_\mathrm{B}\rangle \notag\\
        S_{IJ} &= \langle 0_\mathrm{F}0_\mathrm{B}| [{C}_I, {C}^\dag_J] |0_\mathrm{F}0_\mathrm{B}\rangle \label{eq:eompropagator}
\end{align}
where  $\{ {C}^\dag_I \}$ = $\{ {a}_i, {a}_a, {a}_i {b}^\dag_\nu, {a}_a {b}_\nu\}$, and  $\ket{0}_\mathrm{F}$, $\ket{0}_\mathrm{B}$ are the Fermi and boson vacuums,  
and then the eigenvalue problem
\begin{align}
    \mathbf{H}^{\mathrm{G_0W_0}} \mathbf{R} = \mathbf{R} \mathbf{E}
    \label{eq:GenEigProb}
\end{align}
where $\mathbf{H}^{\mathrm{G_0W_0}} = \mathbf{S}^{-1} \mathbf{H}$, 
yields the $G_0W_0$ quasiparticle energies.

Two choices of $\hat{U}$ have been discussed in Ref.~\citenum{tolle2023exact}: a) One that diagonalizes the Hamiltonian $\hat{H}^\mathrm{B}$ fully, 
and b) One that only block-diagonalizes the Hamiltonian, which can be achieved using the iterative formulation of direct ring unitary coupled-cluster doubles (druCCD).
For b) the druCCD amplitudes 
\begin{align}
    \hat{\sigma}_\mathrm{B} = \frac{1}{2} \sum_{\nu \mu} t_{\nu \mu} \left( \hat{b}^\dagger_\nu \hat{b}^\dagger_\mu - \hat{b}_\nu \hat{b}_\mu \right),
    \label{eq:uCCAmp}
\end{align}
satisfy
\begin{align}
    \bra{0_\mathrm{B}}  [\bar{H}^\mathrm{B}, \hat{b}^\dagger_\nu \hat{b}^\dagger_\mu - \hat{b}_\nu \hat{b}_\mu ] \ket{0_\mathrm{B}}= 0.
\end{align} 
Implemented exactly, both transformations in Eq.~(\ref{eq:CT}) scale as  $\mathcal{O}(N^3_b) = \mathcal{O}(N_o^3 N_v^3)$ ($N_b$ denotes the size of the bosonic space~\cite{tolle2023exact}). In this work, we mainly focus on a), while choice b) is discussed in Appendix~\ref{sec:AppendixA}.

We can improve on this scaling by expanding the bosonic operators in a smaller auxiliary boson (AB) basis. Motivated by studies of density-fitting/resolution-of-the-identity (RI) methods~\cite{vahtras1993integral},  we can choose the size of the auxiliary boson basis $N_\mathrm{AB}$ to scale linearly with the system size.
The AB expansion is the central Ansatz in this work and is given as
\begin{align}
    \hat{b}^\dagger_\nu \approx \sum^{N_\mathrm{AB}}_{Q} C^Q_{\nu} \hat{b}^\dagger_Q, \\
    \hat{b}_\nu \approx \sum^{N_\mathrm{AB}}_{Q} C^Q_{\nu} \hat{b}_Q,
\end{align}
where $Q$ (and capital Latin letters in general) denote AB bosonic indices (and related density fitting indices below).

One can imagine different strategies to construct the AB basis.
For example, we could determine $C^Q_{\nu}$ using information from another correlated calculation, such as from the eigendecomposition of the first order approximation to druCCD amplitudes in Eq.~(\ref{eq:uCCAmp}), which are the direct M\o{}ller-Plesset (MP2) amplitudes \cite{kinoshita2003singular}
\begin{align}
    t_{\nu \mu} &= t_{iajb} = \frac{(ia|jb)}{\epsilon_i + \epsilon_j - \epsilon_a - \epsilon_b} \\
     &= \sum_{Q} C^Q_\nu E_{Q} C^Q_\mu.
     \label{eq:dMP2}
\end{align}
The auxiliary basis vectors $C^Q_\mu$ could then be chosen using a threshold 
$|E_{Q}| > \tau_\mathrm{dMP2}$, to yield a smaller auxiliary basis of size $N_\mathrm{AB}$.

A different approach, and the one we focus on in this work, is to construct $C^Q_\nu$ from a resolution of the identity (RI) Gaussian basis.  
The resolution of the identity (RI) expansion coefficients $R^L_{\nu}$ are defined from the integrals \cite{vahtras1993integral}, i.e.
\begin{align}
 (ia|jb) \approx \sum_{L} R^L_{ia} R^{L}_{jb} \label{eq:RIMOIntegrals},
 \end{align}
with
\begin{align}
R^L_{ia} = \sum_Q(ia|Q)[\mathbf{V}^{-1/2}]_{QP},
\end{align}
where $\mathbf{V}$ denotes the Coulomb metric \cite{eichkorn1995auxiliary}.
We define the auxiliary basis by symmetric orthogonalization of the basis, 
\begin{align}
         C^{Q}_\nu = \sum_{LM} R^L_{\nu} [\mathbf{S}^{-1/2}]_{LM}P^Q_{M},
    \label{eq:AuxBasisExp2}
 \end{align}
where \begin{align}
S_{LM} &= \sum_\nu R^L_\nu R^M_\nu = \sum_{Q} P^Q_L E_{Q} P^Q_M.
    \label{eq:NAF}
\end{align}
The auxiliary basis size may already be smaller than $N_o N_v$, but a further truncated AB basis can be obtained by defining the overlap and its inverse with $E_Q >\tau_\mathrm{RI}$.
This determination of the 
AB basis requires the RI  expansion coefficients in the molecular orbital basis 
(which are obtained with $\mathcal{O}(N^3_\mathrm{orb} N_\mathrm{aux}$), $N_\mathrm{orb}=N_o + N_v$), the construction of $S_{LM}$ with $\mathcal{O}(N_\mathrm{o} N_\mathrm{v} N^2_\mathrm{aux}$), and the determination of the AB basis with $\mathcal{O}(N_\mathrm{AB}N^2_\mathrm{aux})$ cost. We denote the choice of AB expansion by AB(aux), where ``aux'' is the name of the auxiliary RI basis.

\subsection{Practical realization}
\label{sec:practicalrealization}
The Bogoliubov transformation for $\bar{H}^\mathrm{eB}$, expressed in the AB basis, results in (note the electronic part remains unchanged)
\begin{align}
    \bar{H}^\mathrm{eB} &= \hat{H}^e + \sum_{Q} {\Omega}^\mathrm{dRPA}_{Q} \bar{b}^\dagger_Q \bar{b}_Q \nonumber \\
    &+ \sum_{pq,Q} W^{Q}_{pq} \{ \hat{a}^\dagger_p \hat{a}_q\}  (\bar{b}^\dagger_Q + \bar{b}_Q) + E^\mathrm{c}_\mathrm{dRPA},
    \label{eq:ABBogoHam}
\end{align}
with $\Omega^\mathrm{dRPA}_{Q}$ denoting the dRPA excitation energies and ${W}^Q_{pq}$ the transformed electron-boson coupling term
\begin{align}
    {W}^{Q}_{pq} = \sum^{N_\mathrm{AB}}_{R} {V}_{pq,R}\left({\mathbf{X}} + {\mathbf{Y}}\right)_{RQ},
    \label{eq:Wtransformation}
\end{align}
with ${\mathbf{X}}$ and ${\mathbf{Y}}$ being the dRPA excitation vectors.
$E^\mathrm{c}_\mathrm{dRPA}$ is calculated as 
\begin{align}
    E^\mathrm{c}_\mathrm{dRPA} = \frac{1}{2} \mathrm{tr}\{{\pmb{\Omega}} - {\mathbf{A}}\}.
    \label{eq:ABdRPA}
\end{align}

$\bar{H}^\mathrm{eB}$ is constructed in two steps. 
First, the symmetrized dRPA eigenvalue problem is solved in the AB basis
\begin{align}
    \left( {\mathbf{A}} - {\mathbf{B}}\right)^{1/2} \left( {\mathbf{A}} + {\mathbf{B}}\right)  \left( {\mathbf{A}} - {\mathbf{B}}\right)^{1/2} {\mathbf{R}} = {\mathbf{R}} {\pmb{\Omega}}^2.
    \label{eq:Symmetrized}
\end{align}
The dRPA excitation energies are obtained from the square root of $~{\pmb{\Omega}}^2$.
The excitation vectors $\left(\mathbf{X} + \mathbf{Y}\right)$ are then reconstructed using $\mathbf{R}$ (see e.g. Ref.~\citenum{furche2001density} for the explicit expressions).
These steps [determination of $\pmb{\Omega}$ and $\left(\mathbf{X} + \mathbf{Y}\right)$] scale as $\mathcal{O}(N^3_\mathrm{AB})$.
Next, the eigenvectors are used to transform the electron-boson coupling [Eq.~(\ref{eq:Wtransformation})] which scales as $\mathcal{O}(N^2_\mathrm{orb} N^2_\mathrm{AB})$ ($N_\mathrm{orb}$ denotes the number of electronic orbitals). This results in an overall quartic scaling of the Hamiltonian transformation. 

As discussed below, $N_\mathrm{AB}$ is typically larger than the size of standard RI basis, $N_\mathrm{aux}$. Thus, the prefactor of the Hamiltonian transformation can be further reduced by using a smaller RI basis to represent the integrals,
\begin{align}
    W^{P}_{pq} = \sum^\mathrm{N_\mathrm{AB}}_{Q} \sum^{N_\mathrm{aux}}_{{L}} \sum_\nu
    R_{pq}^L  R_{\nu}^L C_{\nu}^Q \left(\mathbf{X} + \mathbf{Y} \right)_{QP},
    \label{eq:WtransformationRI}    
\end{align}
where the contractions can be performed with $\mathcal{O}(N_\mathrm{aux}N^2_\mathrm{AB}) + \mathcal{O}(N^2_\mathrm{orb}N_\mathrm{aux} N_\mathrm{AB})$ cost, which is a savings if $N_\mathrm{AB} >  N_\mathrm{aux}$.

Using $\bar{H}^\mathrm{eB}$, we obtain selected quasiparticle energies from Eq.~(\ref{eq:GenEigProb}) using 
the Davidson iterative diagonalization procedure. 
The  $\sigma$-vector equations for an excitation $n$ are given as (the excitation vector is denoted as $r$)
\begin{align}
    \sigma^n_i &= \sum_{j}f_{ij}r^n_j + \sum_{a} f_{ia} r^n_{a} \nonumber \\
    &+ \sum_{Q j} W^Q_{ji} r^n_{j Q}  + \sum_{Q a} W^Q_{i a} r^n_{Q a} \\
    \label{eq:1}
    \sigma^n_a &= \sum_{b}f_{ab}r^n_b + \sum_{i} f_{ia} r^n_{i} \nonumber \\
    &+ \sum_{Q i} W^Q_{i a} r^n_{i Q}  +  \sum_{Q b} W^Q_{a b} r^n_{Q b} \\
    \sigma^n_{iP} &= \sum_j f_{ij} r^n_{j P} + \sum_{j} W^Q_{ji} r^n_j \nonumber \\
    &+ \sum_{a} W^Q_{ia} r^n_a + \sum_{Q} \bar{A}_{PQ} r^n_{iQ} \\
    \sigma^n_{aP} &= \sum_b f_{ab} r^n_{b P} + \sum_{i} W^Q_{ia} r^n_i \nonumber \\
    &+ \sum_{b} W^Q_{ab} r^n_b + \sum_{Q} \bar{A}_{PQ} r^n_{aQ}.
    \label{eq:4}
\end{align}
where the scaling for the matrix-vector products (for each root $n$) is $\mathcal{O}(N_\mathrm{orb}^2 N_\mathrm{AB})$/$\mathcal{O}(N_\mathrm{orb} N^2_\mathrm{AB})$ (depending on $N_\mathrm{orb}> N_\mathrm{AB}$ or $N_\mathrm{AB} > N_\mathrm{orb}$), i.e. cubic with system size [note that $\bar{A}_{PQ}$ is diagonal when using the transformed Hamiltonian as defined in Eq.~(\ref{eq:ABBogoHam})].
One complication arises from the fact that the valence quasiparticle energies (IP and EA) do not correspond to the lowest eigenvalues of $\mathbf{H}^\mathrm{G_0W_0}$.
In practice, we calculate valence quasiparticle energies using a root following procedure starting from guess vectors associated with valence orbitals.

The above procedure determines the G$_0$W$_0$ eigenvalues without the commonly used diagonal approximation where the self-energy is assumed diagonal \cite{kaplan2015off}.
The diagonal approximation can be restored by appropriately restricting the summations of Eqs.~(\ref{eq:1})-(\ref{eq:4}).
This also fixes one of the $pq$ orbital indices of $W^Q_{pq}$ when considering a specific excitation \cite{bintrim2021full}.

\section{Computational details}
We implemented the above procedure, which we term the AB-G$_0$W$_0$ method, in $\textsc{PySCF}$ \cite{sun2020recent}.
Reference orbitals are Hartree--Fock (HF) orbitals if not specified otherwise.
EOM Davidson iterations were truncated after the residual was smaller than $10^{-6}$ Ha. 
Except when otherwise specified, in all calculations we used the RI approximation for the two-electron integrals, using the RI auxiliary basis associated with the orbital  basis set \cite{weigend1998ri}.
The AB basis was constructed either from predefined RI auxiliary basis sets, or more flexible even-tempered (ET) \cite{raffenetti1973even} auxiliary basis sets using the construction scheme implemented in \textsc{PySCF}. These ET bases are defined as follows:
For a specific element, the exponents of the auxiliary basis functions, given an angular momentum $l$, are determined as
\begin{align}
    \alpha^l_i = \beta \alpha^l_{i-1};\quad i = 1 \dots N^l,
\end{align}
with $\beta$ being a user-defined input parameter, and $N^l$ is determined from
\begin{align}
    N^l = \Bigg\lceil \frac{\log(\frac{\alpha^l_\mathrm{aux,max} + \alpha^l_\mathrm{aux,min}}{\alpha^l_\mathrm{aux,min}})}{\log(\beta)} \Bigg\rceil ,
\end{align}
where $\alpha^l_\mathrm{aux,max}$ and $\alpha^l_\mathrm{aux,min}$ denote effective exponents of the auxiliary basis for a given $l$ ($\alpha^l_{0} = \alpha^l_\mathrm{aux,min}$), and $\lceil \dots \rceil$ denotes the ceil function.
The angular momenta of the auxiliary basis are chosen to be $l = \{0,\dots,2\tilde{l}\}$, with $\tilde{l}$ denoting the angular momenta of the `parent basis', i.e.~def2-TZVP.
The maximum and minimum exponents of the auxiliary basis for a given angular moment $l$ are determined as
\begin{align}
    \alpha^l_\mathrm{aux,max} = 2\times\mathrm{max}\left\{ \sqrt{\alpha^{\tilde{l}}_\mathrm{parent,max} \alpha^{\tilde{l}'}_\mathrm{parent,max}} \vert \tilde{l} + \tilde{l}' = l\right\}
\end{align}
and 
\begin{align}
    \alpha^l_\mathrm{aux,min} = \mathrm{min}\left\{ \sqrt{\alpha^{\tilde{l}}_\mathrm{parent,min} \alpha^{\tilde{l}'}_\mathrm{parent,min}} \vert \tilde{l} + \tilde{l}' = l\right\},
\end{align}
with $\alpha^{\tilde{l}}_\mathrm{parent,min}/\alpha^{\tilde{l}}_\mathrm{parent,max}$ denoting the maximum and minimum exponents in the `parent basis'.
\section{Applications}

\subsection{Small molecule benchmarks}

As a first benchmark of the AB ansatz, we consider the error of the ground-state dRPA correlation energy in the AB basis. We compute $E_\mathrm{dRPA}$ [Eq.~(\ref{eq:ABdRPA})] using both the full bosonic basis and the AB basis to obtain $\pmb{\Omega}$ and $\mathbf{A}$.
We consider a set of 12 molecular- and atomic systems (helium, neon, hydrogen, fluorine, silane, carbon monoxide, water, beryllium monoxide, magnesium monoxide, formaldehyde, methane, sulfur dioxide) with structures taken from the GW100 benchmark set \cite{van2015gw}. For all systems, we use the def2-TZVP orbital basis and integrals in the RI approximation using the corresponding def2-TZVP-RI basis. Over this class of systems, the MAE of the dRPA correlation energy in the full boson basis, but using the RI integral approximation instead of the exact integrals, is 4.4 meV.

\begin{figure}
    \centering
    \includegraphics[width=1.0\linewidth]{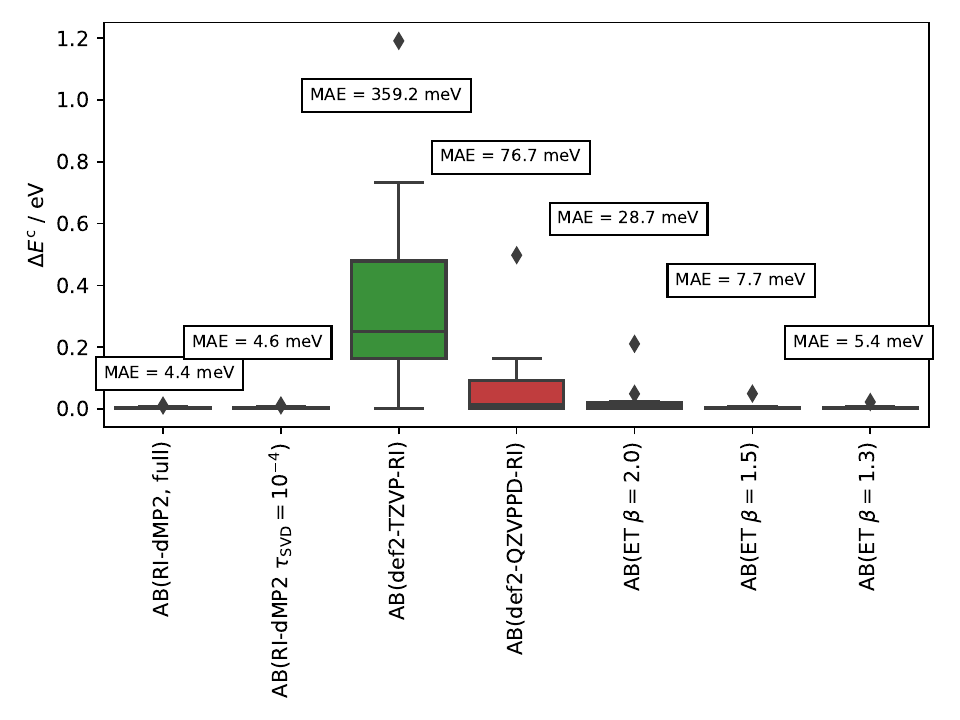}
    \caption{Box-plot representation of the error (in eV) in the dRPA energies of 12 small atomic and molecular systems computed using the auxiliary boson  (AB) ansatz (relative to the exact dRPA correlation energies in the def2-TZVP basis, see text for more details).
    The AB basis is specified in parentheses. The mean absolute error (MAE) in meV is shown as an inset.}
    \label{fig:Ecbenchmark}
\end{figure}

The errors of the AB dRPA energies are displayed in  Fig.~\ref{fig:Ecbenchmark}.
We see that 
AB(def2-TZVP-RI), which uses the same auxiliary basis to represent the integrals and for the AB expansion, yields a large error with a MAE of 359.2 meV.
This highlights the need to use different auxiliary basis sets for the integrals and for the AB ansatz in Eq.~(\ref{eq:AuxBasisExp2}).
Using the larger def2-QZVPPD-RI auxiliary boson basis reduces the error significantly. Using the ET basis, we approach completeness in the AB basis by reducing the parameter $\beta$, and we see systematic convergence of the correlation energy.  
 For $\beta = 1.5$ the MAE is similar to the intrinsic error of using the RI approximation to the integrals discussed above.

We note that the exact dRPA correlation energy in the given orbital basis when computed with or without the RI approximation to the integrals, is a lower bound to the dRPA energies within the AB approximation in combination with the RI approximation to the integrals.
For an additional discussion and proof, we refer to Appendix \ref{sec:AppendixB}.

\begin{figure}
    \centering
    \includegraphics[width=1.0\linewidth]{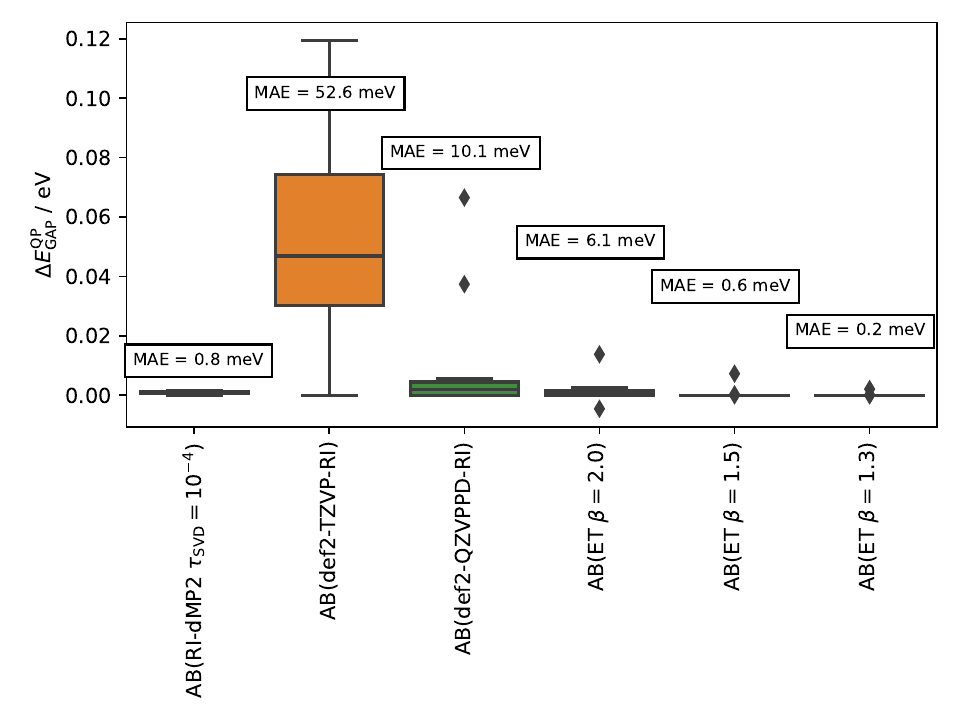}
    \caption{Box-plot representation of the error (in eV) of the AB-G$_0$W$_0$ quasiparticle gap relative to the  to CD-G$_0$W$_0$ quasiparticle gap. The AB basis is specified in parentheses. The mean absolute error (MAE) in meV is shown as the inset.}
    \label{fig:Gapbenchmark}
\end{figure}

Next, we evaluate the accuracy of the AB ansatz for the quasiparticle gap in the same systems.
We obtained reference quasiparticle energies using the fully analytic G$_0$W$_0$ implementation in \textsc{PySCF}.
The \textsc{PySCF} fully analytic G$_0$W$_0$ implementation uses the
diagonal approximation to evaluate the self-energy \cite{bruneval2012ionization,van2013gw}.
Consequently, all AB-G$_0$W$_0$ calculations were also performed in the diagonal approximation by truncating the sums in Eqs.~(\ref{eq:1}-\ref{eq:4}) (see also Ref.~\citenum{bintrim2021full}). 
The box plots for the errors and the MAE are shown in Fig.~\ref{fig:Gapbenchmark}.
Similar trends are observed as for the correlation energies, 
however, the overall MAE is smaller.

\subsection{Scaling}

\begin{figure*}
    \centering
    \includegraphics[width=1.0\linewidth]{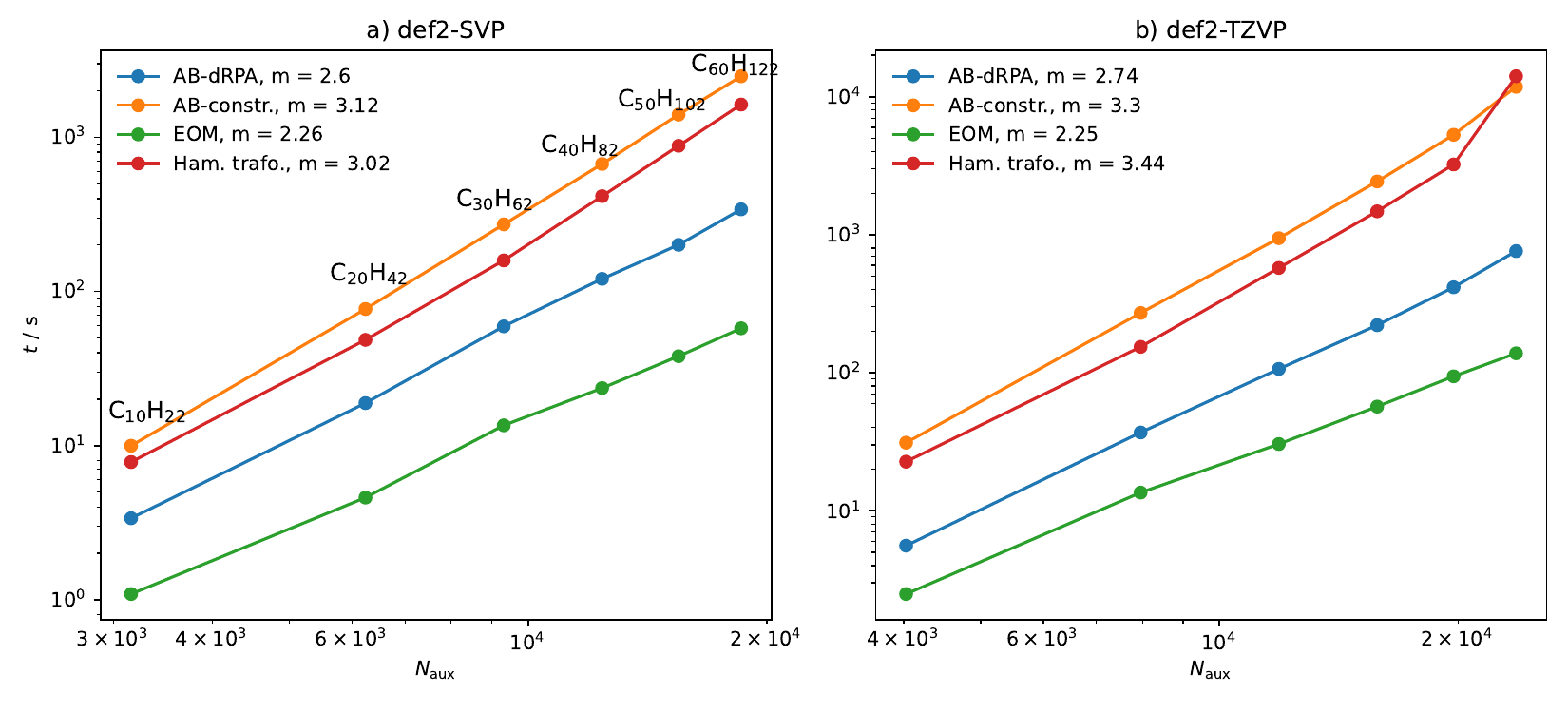}
    \caption{Wall-clock timings for the AB-dRPA, AB basis construction (AB-constr.), the construction of $\mathbf{A},\mathbf{B}$ in the AB basis and the quartic $\mathcal{O}(N^2_\mathrm{orb}N_\mathrm{aux} N_\mathrm{AB})$-step in the construction of $W^P_{pq}$ (denoted as Ham. trafo.), and the EOM iterations (denoted EOM) as a function of the size of auxiliary basis in linear alkanes C$_n$H$_{2n+2}$ ($n = 10 - 60$) for the a) def2-SVP and b) def2-TZVP orbital basis sets. The slope of the timings is denoted as $m$ in the legend.}
    \label{fig:TimingsLinearAlcane}
\end{figure*}

Having established the accuracy of the AB-G$_0$W$_0$ method, we now test the computational scaling of the implementation. We study the linear alkanes C$_n$H$_{2n+2}$ for $n = 10 - 60$ in the def2-SVP and def2-TZVP basis and compute the HOMO and the LUMO quasiparticle energies within the diagonal approximation.
For the auxiliary boson basis, we use an ET auxiliary basis with $\beta=1.5$.
The timings for the diagonalization of $\hat{H}^B$ [Eq.~(\ref{eq:Symmetrized}), denoted AB-dRPA, including the cubic contraction of Eq.~(\ref{eq:WtransformationRI})], the construction of the AB basis [denoted AB-constr., Eqs.~(\ref{eq:RIMOIntegrals}-\ref{eq:NAF})], the construction of  the initial $\mathbf{A},\mathbf{B}$ in the AB basis and the quartic $\mathcal{O}(N^2_\mathrm{orb}N_\mathrm{aux} N_\mathrm{AB})$-step in the construction of $W^P_{pq}$ (denoted as Ham. trafo.), and the EOM iterations (denoted EOM) for the def2-SVP and def2-TZVP basis are shown in the double logarithmic plot in Fig.~\ref{fig:TimingsLinearAlcane}. 

The AB-dPRA and EOM parts of the calculation have a slope between $2-3$ reflecting the theoretical cubic scaling, while
the construction of the AB basis and  the Hamiltonian transformations have a slope between $3-4$, reflecting their theoretical quartic scaling, respectively.

\begin{figure}
    \centering
    \includegraphics[width=1.05\linewidth]{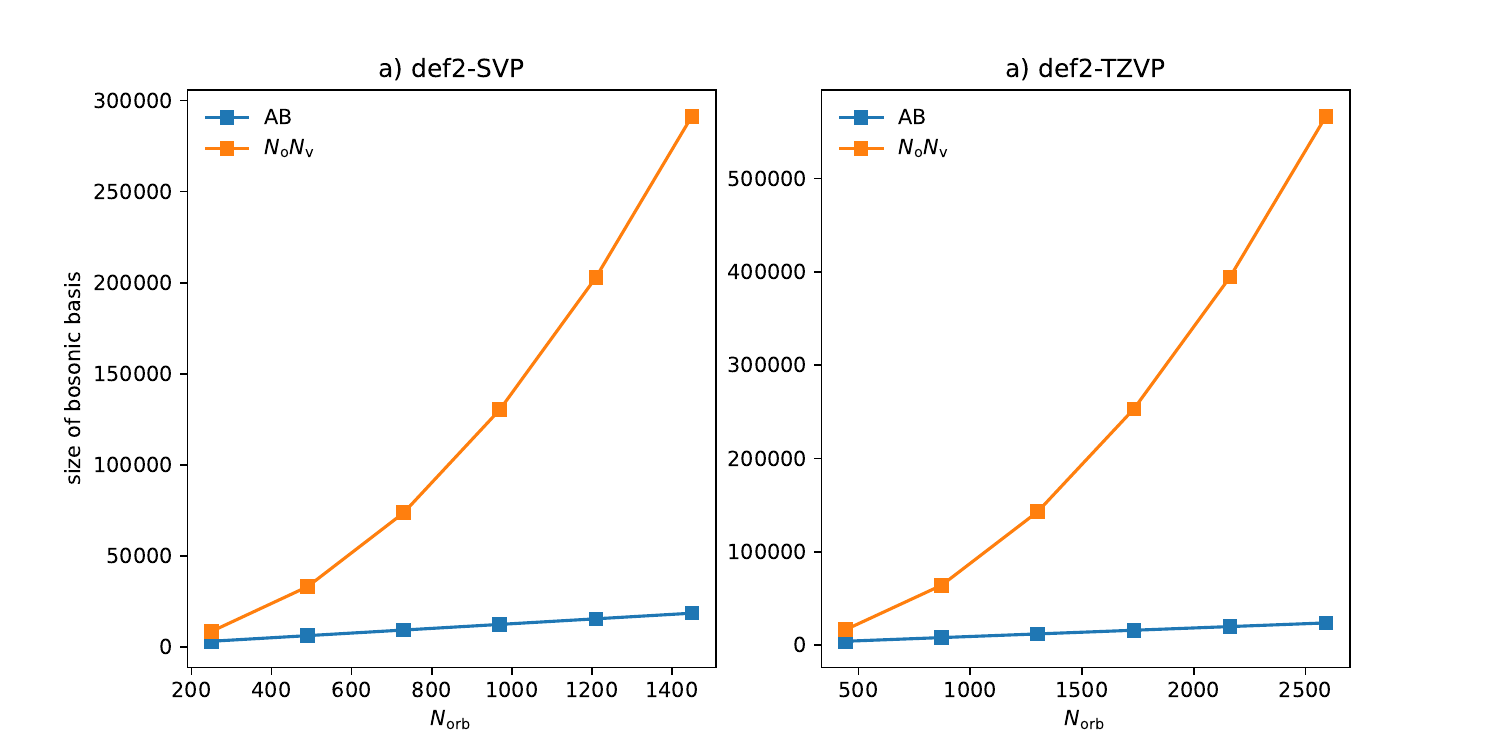}
    \caption{Size of full boson ($N_oN_v$) and auxiliary boson basis [AB(ET-$\beta=1.5$)] for the a) def2-SVP and b) def2-TZVP basis set for linear alkanes C$_n$H$_{2n+2}$ ($n = 10 - 60$).}
    \label{fig:BasisSize}
\end{figure}

The size of the full bosonic basis ($N_oN_v$) and the AB basis [AB(ET-$\beta=1.5$)] as a function of system size is displayed in Fig.~\ref{fig:BasisSize}.
While the bosonic basis grows quadratically, the AB basis grows linearly by construction, which is the origin of the improved scaling of the AB-G$_0$W$_0$ method.

For $n=10-30$ (def2-SVP) and $n=10-20$ (def2-TZVP) we also show the error in the HOMO and LUMO quasiparticle energies using the AB basis relative to the  the full bosonic basis in 
Tab.~\ref{tab:LinearAlcaneDeviation}.
In the case of the def2-SVP basis, the maximum deviation for the AB basis, using an ET auxiliary basis ($\beta=1.5$), is $4$~meV and $2$~meV for the HOMO and LUMO energies, respectively. 
For the def2-TZVP basis, the maximum  deviation for the HOMO is $20$~meV and $9$~meV for the LUMO.
These deviations are all significantly below the intrinsic error of the G$_0$W$_0$ method.
We also see that the error is almost independent of the system size. 
\begin{table}[!t]
    \centering
    \caption{Absolute deviation in the HOMO/LUMO quasiparticle energies in eV computed using the AB basis (ET-$\beta=1.5$) relative to the full bosonic basis for a linear alkane chain C$_n$H$_{2n+2}$ ($n = 10 - 30$) using the def2-SVP/def2-TZVP orbital basis.}
    \label{tab:LinearAlcaneDeviation}
    \begin{tabular}{l r r r r}
        \hline
        & \multicolumn{2}{c}{def2-SVP} & \multicolumn{2}{c}{def2-TZVP}\\
        \hline
        $n$ & \multicolumn{1}{c}{HOMO / eV} & \multicolumn{1}{c}{LUMO / eV} & \multicolumn{1}{c}{HOMO / eV} & \multicolumn{1}{c}{LUMO / eV} \\
        \hline
        10 & 0.003 & 0.001 & 0.016 & 0.007 \\
        20 & 0.004 & 0.002 & 0.020 & 0.009 \\
        30 & 0.004 & 0.002 & - &  - \\
        \hline
    \end{tabular}
\end{table}
\subsection{Core level binding energies}
Next we demonstrate the computation of core-level binding energies (CBE) using AB-G$_0$W$_0$.
Core-level binding energies are a challenge for many G$_0$W$_0$ implementations due to the steep computational scaling  of the ``fully-analytic'' and CD-G$_0$W$_0$ approaches with the number of electrons, or the inaccuracy of the AC-G$_0$W$_0$ approach \cite{golze2018core}.
Thus improved algorithms for core-level binding energies within the G$_0$W$_0$ approximation is an ongoing field of research, c.f.~Refs.~\citenum{duchemin2020robust,panades2023accelerating}.

We first benchmark the AB-G$_0$W$_0$ method for 10 CBEs of small molecules. 
Structures, experimental values, and CD-G$_0$W$_0$ quasiparticle energies starting from PBEh (with 45 \% exact exchange) mean-field orbitals in the cc-pVTZ basis, were taken from Ref.~\citenum{golze2020accurate}.
The relative errors, compared to experiment, for CD-G$_0$W$_0$ and G$_0$W$_0$ using the full boson basis and various auxiliary boson basis (and with and without the diagonal approximation) are presented 
in Tab.~\ref{tab:CBEBenchmark}. We also consider the effect of a scalar relativistic correction using the ``exact two component'' (X2C) Hamiltonian.
In the case of benzene, the average of the lowest six quasiparticle energies is shown. 

CBEs from Ref.~\citenum{golze2020accurate} compared to our G$_0$W$_0$ implementation with the full bosonic basis lead to 
almost identical core-level binding energies.
Furthermore, the CBEs with and without the diagonal approximation are almost identical, illustrating that the off-diagonal self-energy elements are small for the core orbitals. The relativistic contribution is significant and reduces the MAE (relative to experiment) of the full boson basis G$_0$W$_0$  from 0.75 eV to 0.55 eV.

The auxiliary boson basis results systematically converge to the full boson basis results as the ET basis size increases. 
With the exception of benzene, the deviation for ET-$\beta=1.3$ from the full boson basis is smaller than 0.01 eV. Compared to the intrinsic error of the G$_0$W$_0$ method (as measured by the error of the full bosonic basis result), even using the smallest ET basis (large $\beta$) already yields accurate results.

\begin{table*}
    \caption{Core-level binding energies for various molecular systems relative to experiment [starting orbitals PBEh with 45\% exact exchange, orbital basis: cc-pVTZ]. Diag. indicates the diagonal approximation to the self-energy. Full represents the full bosonic basis, ET-$\beta$ is a AB calculation with an even-tempered basis. X2C: orbitals and integrals from an exact-two-component mean-field calculation.}
    \begin{tabular}{lcccccccc}
    \hline
    molecule (core level) &  exp.$^{a}$ &  Golze \textit{et al.}$^{b}$ & Full diag. & Full &  ET-$\beta=2$ diag. &  ET-$\beta=1.5$ diag. &  ET-$\beta=1.3$ diag. &  Full (X2C) \\
    \hline
    benzene (C1s) &   290.38 & $-$0.49 & $-$0.50 & $-$0.49 & $-$0.01 & $-$0.20 & $-$0.32 & $-$0.38     \\
    CO$_2$ (O1s)     &  541.32 & $-$1.10 & $-$1.10 & $-$1.10 & $-$0.99 & $-$1.24 & $-$1.11 & $-$0.75   \\
    CO$_2$ (C1s)     &   297.70 & $-$0.31 & $-$0.31 & $-$0.31 & $-$0.20 & $-$0.28 & $-$0.31 & $-$0.22 \\
    CO (O1s)      &  542.10 & $-$0.95 & $-$0.94 & $-$0.94 & $-$0.95 & $-$0.94 & $-$0.94 & $-$0.59 \\
    CO (C1s)      &   296.23 & $-$0.75 & $-$0.75 & $-$0.75 & $-$0.73 & $-$0.75 & $-$0.75 & $-$0.66 \\
    methane (C1s)  &  290.84 & $-$0.64 & $-$0.64 & $-$0.64 & $-$0.64 & $-$0.64 & $-$0.64 & $-$0.56  \\
    ethane (C1s) & 290.71 & $-$0.53 & $-$0.53 & $-$0.53 & $-$0.43 & $-$0.53 & $-$0.53 & $-$0.45 \\
    formaldehyde (O1s)  & 539.33 & $-$1.18 & $-$1.18 & $-$1.18 & $-$1.11 & $-$1.18 & $-$1.18 & $-$0.83   \\
    formaldehyde (C1s)    &  294.38 & $-$0.35 & $-$0.35 & $-$0.35 & $-$0.33 & $-$0.35 & $-$0.35 & $-$0.27  \\
    water (O1s)   & 539.70 & $-$1.15 & $-$1.15 & $-$1.15 & $-$1.15 & $-$1.15 & $-$1.15 & $-$0.80   \\
    \hline
    MAE & $-$ & 0.75 & 0.75 & 0.74 & 0.65 & 0.73 & 0.73 & 0.55\\
    \hline
    \end{tabular}
    \begin{tiny}
        a)/b) values taken from Ref.~\citenum{golze2020accurate}\\
    \end{tiny}
    \label{tab:CBEBenchmark}
\end{table*}

\begin{figure}
    \centering
    \includegraphics[width=1.0\linewidth]{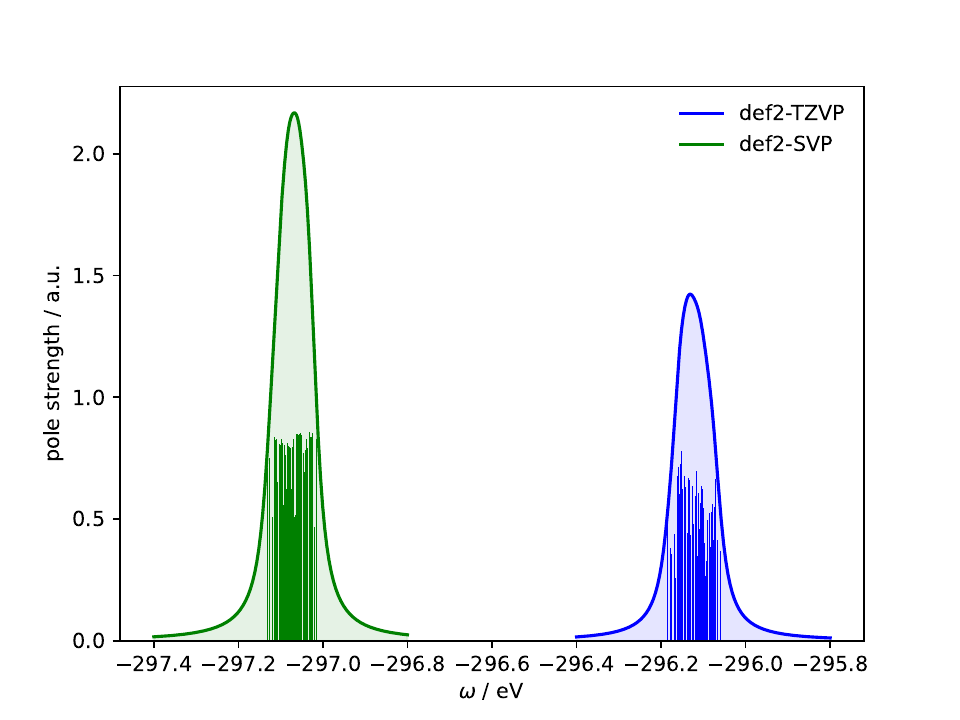}
    \caption{Spectral function for the sixty carbon 1s core-states of C$_{60}$. The curves use a Lorentzian ($\eta = 0.02$ eV) broadening of the pole strengths (shown by bars).}
    \label{fig:C60}
\end{figure}

 We further demonstrate the practical applicability of our approach to determine CBEs in a molecule with a large number of electrons, namely C$_{60}$. When there are a large number of electrons, the reference CD-G$_0$W$_0$ method becomes expensive because of the $\mathcal{O}(N_o N_v N_\mathrm{val} N_\mathrm{aux}^2)$ scaling~\cite{zhu2021all}. 
 The spectral function for the core-level binding energies associated with all 60 C1s-states (within the diagonal approximation) for the def2-SVP and the def2-TZVP basis sets, using the ET auxiliary basis with $\beta=1.5$, is displayed in Fig.~\ref{fig:C60} (starting from Hartree--Fock orbitals), using a  Lorentzian broadening ($\eta = 0.02$ eV). Although a fair comparison of timings is complicated by implementation details, we note that the AB-G$_0$W$_0$ calculation for the lowest quasiparticle energy was approximately 12 times faster than the CD-G$_0$W$_0$ implementation in \textsc{PySCF} (using 60 imaginary frequency points).

\section{Conclusion and Outlook}

In summary, we have demonstrated a practical method for the frequency-free determination of G$_0$W$_0$ quasiparticle energies, by using an auxiliary-boson (AB) ansatz within the equation-of-motion formulation of the G$_0$W$_0$ quasiparticle problem.  

The new method is named AB-G$_0$W$_0$. We show its  applicability by computing valence and core quasiparticle energies in a variety of molecular systems including ones with more than 180 atoms. 
Because the new approach avoids frequency integration, it does not face particular problems in treating the core quasiparticle energies.

The main cost of the AB-G$_0$W$_0$ method arises from the construction of the AB basis and the transformation of the electron-boson Hamiltonian into this representation. The resulting quartic cost is closely tied to the use of the resolution-of-the-identity basis. The scaling can therefore be reduced by using improvements to the resolution-of-the-identity formalism, for example, through the ideas of separable density-fitting \cite{lu2015compression,duchemin2019separable}, whereby the scaling of the AB basis construction and transformation are improved to cubic. These ideas will be explored in future work. 

\section*{Acknowledgments}
This work was supported by the US Department of Energy, Office of Science, via Award no. DE-SC0018140.
JT acknowledges funding through a postdoctoral research fellowship from the Deutsche Forschungsgemeinschaft (DFG, German Research Foundation) – 495279997. 

\section{Appendix}
\label{sec:Appendix}
\subsection{Appendix A}
\label{sec:AppendixA}
Within the AB expansion (Sec.~\ref{sec:Theory}), the druCCD amplitudes $t_{PQ}$ can be computed using the machinery of canonical-transformation theory (CT), developed in molecular quantum chemistry, c.f.~Refs.~\citenum{neuscamman2010review,yanai2006canonical}.
Starting from an initial (denoted by the $(0)$ superscript) set of amplitudes $t^{(0)}_{PQ}$ (resulting in $\hat{\sigma}^{(0)}_B$), $\bar{H}^\mathrm{eB}$ is constructed iteratively
\begin{align}
    \bar{H}^\mathrm{eB,(n+1)} = [\bar{H}^\mathrm{eB,(n)},\hat{\sigma}^{(0)}_B],
    \label{eq:HamConstr}
\end{align}
so that
\begin{align}
   \bar{H}^\mathrm{eB} = \sum_{n=0} \frac{\bar{H}^\mathrm{eB,(n)}}{n!}.  
\end{align}
In practice, the sum is truncated after $\frac{\bar{H}^\mathrm{eB,(n)}}{n!} < 10^{-9}$.
Note that the evaluation of the BCH expansion is numerically exact (up to the truncation threshold) because  $\bar{H}^\mathrm{eB}$ and $\hat{\sigma}_B$  contain quadratic bosonic contributions only.
The commutator expressions for the evaluation of Eq.~(\ref{eq:HamConstr}) are given as 
\begin{align}
[\bar{H}^\mathrm{eB,(n)},\hat{\sigma}^0_B] &= \hat{H}^\mathrm{e} \nonumber \\
    &+ \sum_{pq,Q} \underbrace{{V}^{(n)}_{pq Q} {t}^{(0)}_{QR}}_{{V}^{(n+1)}_{pq R}} \{ \hat{a}^\dagger_p \hat{a}_q\}  (\hat{b}^\dagger_R + \hat{b}_R) \nonumber \\ 
    &+ \frac{1}{2} \sum_{Q} \underbrace{({A}^{(n)}_{P Q}{t}^{(0)}_{QR} + {t}^{(0)}_{P Q} {A}^{(n)}_{QR})}_{B^{(n+1)}_{PR}} (\hat{b}^\dagger_P \hat{b}^\dagger_R + \hat{b}_P \hat{b}_R) \nonumber \\ 
    &+\sum_{Q} \underbrace{({B}^{(n)}_{P Q}{t}^{(0)}_{QR} + {t}^{(0)}_{P Q} {B}^{(n)}_{QR})}_{A^{(n+1)}_{PR}} \hat{b}^\dagger_P \hat{b}_R \nonumber \\
    &+ H^{(n)}_0.
    \label{eq:hebHam}
\end{align}
The correlation energy from the initial amplitudes ($\mathbf{t}^{(0)}$) is 
\begin{align}
    E^{c}_\mathrm{druCCD} = \sum_n \frac{1}{n!} H^{(n)}_0 = \sum_n \frac{1}{n!} \mathrm{tr}\{ {\mathbf{B}}^{(n)} {\mathbf{t}}^{(0)}\}.
    \label{eq:correlationEnergy}
\end{align}
In practice, we evaluate the correlation energy in the AB basis as 
\begin{align}
    E^{c}_\mathrm{druCCD} = \frac{1}{2} \mathrm{tr}\{ \bar{\mathbf{A}} - {\mathbf{A}} \},
    \label{eq:energyEval}
\end{align}
with $\bar{\mathbf{A}}$ and ${\mathbf{A}}$ denoting the transformed and the not-transformed $\mathbf{A}$ matrix appearing in $\bar{H}^\mathrm{B}$ and $\hat{H}^\mathrm{B}$, respectively.
The advantage of using Eq.~(\ref{eq:energyEval}) for the energy evaluation is that it has to be evaluated only once, after $\bar{H}^\mathrm{eB}$ is constructed.
$E^{\mathrm{c}}_{\mathrm{druCCD}}$ is identical to the direct random-phase-approximation (AB-dRPA) correlation energy [Eq.~(\ref{eq:ABdRPA})], $E^{\mathrm{c}}_\mathrm{dRPA}$ in the same AB basis. 
One can see that the determination of the electron-boson coupling term has the dominant cost, scaling as $\mathcal{O}(N^2_\mathrm{orb} N^2_\mathrm{AB})$.
In practice, we minimize the prefactor for the construction of $\bar{H}^\mathrm{eB}$ by making use of the RI approximation, ${V}^n_{pq Q} = \sum^{N_\mathrm{aux}}_N R^N_{pq} {V}^n_{N Q}$, so that
\begin{align}
    &\sum_{pq,Q} V^{(n)}_{pq Q} t^{(0)}_{QR} \{ \hat{a}^\dagger_p \hat{a}_q\}  (\hat{b}^\dagger_R + \hat{b}_R) \nonumber \\
    &= \sum_{pq,Q} \sum^{N_\mathrm{aux}}_N R_{pqN} ({V}^{(n)}_{N Q} t^{(0)}_{QR}) \{ \hat{a}^\dagger_p \hat{a}_q\}  (\hat{b}^\dagger_R + \hat{b}_R)
    \label{eq:HalfTransform}
\end{align}
and the contraction of $R^N_{pq}$ with $(\sum^{N_\mathrm{AB}}_Q {V}^{(n)}_{N Q} t^{(0)}_{QR})_{NR}$, which scales  as $\mathcal{O}(N^2_\mathrm{orb} N_\mathrm{AB} N_\mathrm{aux}$), has to be performed only once, while the scaling for the iterative step is reduced to $\mathcal{O}(N_\mathrm{aux} N^2_\mathrm{AB})$. All remaining steps scale as $\mathcal{O}(N_{AB}^3)$.

As an example of the amplitude iteration, we consider the correction to the amplitudes for the first iteration, $\Delta^{(1)}_B$.
This is evaluated by solving
\begin{align}
    &\bra{0_\mathrm{B}0_\mathrm{F}} [\bar{H}^\mathrm{eB}, \Delta^{(1)}_{PQ} ( \hat{b}^\dagger_P \hat{b}^\dagger_Q - \hat{b}_P \hat{b}_Q) ] \ket{0_\mathrm{B}0_\mathrm{F}} \nonumber \\ 
    &= -\bra{0_\mathrm{B}0_\mathrm{F}} [\bar{H}^\mathrm{eB}, \hat{b}^\dagger_P \hat{b}^\dagger_Q - \hat{b}_P \hat{b}_Q ] \ket{0_\mathrm{B}0_\mathrm{F}},
\end{align}
using the Krylov subspace method.
The updated amplitudes ${t}^{(1)}_{PQ}$ are then obtained as
\begin{align}
    {t}^{(1)}_{PQ}= {t}^{(0)}_{PQ} + \Delta^{(1)}_{PQ}.
\end{align}
Based on ${t}^{(1)}_{PQ}$ a new $\bar{H}^\mathrm{eB}$ is constructed from which a new correction to the amplitudes is obtained. The procedure is repeated until the change in the amplitudes and the correlation energy is below the desired thresholds.
\subsection{Appendix B}
\label{sec:AppendixB}
In the following, we show that the AB-dRPA correlation energy ($E^{\mathrm{AB},c}_\mathrm{dRPA}$) is always above or equal to $E^c_\mathrm{dRPA}$ in the full boson basis,
\begin{align}
    E^{\mathrm{AB},c}_\mathrm{dRPA} &\geq E^c_\mathrm{dRPA} \nonumber \\
    \frac{1}{2} \mathrm{tr}\{{\pmb{\Omega}}^\mathrm{AB} - {\mathbf{A}}^\mathrm{AB}\} &\geq \frac{1}{2} \mathrm{tr}\{{\pmb{\Omega}} - \mathbf{A}\}.
    \label{eq:dRPAEnergyBound}
\end{align}
where in this section we explicitly denote  quantities expanded in the AB basis with the AB superscript.
In the case of dRPA, it can be shown that $E^c_\mathrm{dRPA}$ calculated within the RI approximation to the integrals represents a variational upper bound to $E^c_\mathrm{dRPA}$ without the RI approximation, assuming fixed dRPA eigenvectors and mean-field orbitals~\cite{eshuis2010fast}.

Here we also assume the same initial orbitals are used in the AB basis and in the full boson basis dRPA calculation. 
Expressing the correlation energy in terms of the energy expression for the the druCCD case [Eq.~(\ref{eq:energyEval})], we find for the energy difference
\begin{align}
    \Delta E^c &= E^c_\mathrm{dRPA} - E^{\mathrm{AB},c}_\mathrm{dRPA} \nonumber \\
    &= \frac{1}{2} \mathrm{tr}\{ \bar{\mathbf{A}} - \mathbf{A}\} - \frac{1}{2} \mathrm{tr}\{ \bar{\mathbf{A}}^\mathrm{AB} - \mathbf{A}^\mathrm{AB}\},
    \label{eq:Difference}
\end{align}
with [Eq.~(\ref{eq:correlationEnergy})]
\begin{align}
    E^c_\mathrm{dRPA} =  \mathrm{tr}\{ \underbrace{\sum_n \frac{1}{n!} \mathbf{B}^{(n)} \mathbf{t}}_{\bar{\mathbf{A}}^c}\}, \label{eq:Acdef}
\end{align}
and 
\begin{align}
    E^{\mathrm{AB},c}_\mathrm{dRPA} = \mathrm{tr}\{ \underbrace{\sum_n \frac{1}{n!} \mathbf{B}^{\mathrm{AB},(n)} \mathbf{t}^{\mathrm{AB}}}_{\bar{\mathbf{A}}^{\mathrm{AB},c}} \}.
\end{align}
So that Eq.~(\ref{eq:Difference}) is rewritten as
\begin{align}
    \Delta E^c &= \frac{1}{2} \mathrm{tr}\{ \bar{\mathbf{A}}^c\} - \frac{1}{2} \mathrm{tr}\{ \bar{\mathbf{A}}^{\mathrm{AB},c}\}.
    \label{eq:corrbasis}
\end{align}
Note that
\begin{align}
    \bar{\mathbf{A}}^{\mathrm{AB},c} = \mathbf{C}^\mathrm{AB} \bar{\mathbf{A}}^{c} \mathbf{C}^{\mathrm{AB},T},
\end{align}
where $\mathbf{C}^\mathrm{AB}$ denotes the auxiliary basis transformation coefficients [Eq.~(\ref{eq:AuxBasisExp2})].
Due to the invariance of the trace under cyclic permutation, Eq.~(\ref{eq:corrbasis}) is rewritten
\begin{align}
    \Delta E^c &= \frac{1}{2} \mathrm{tr}\{ \bar{\mathbf{A}}^c  - \bar{\mathbf{A}}^{c} \mathbf{C}^\mathrm{AB} \mathbf{C}^{\mathrm{AB},T}  \},
\end{align}
with $\mathbf{C}^\mathrm{AB} \mathbf{C}^{\mathrm{AB},T}$ is the projector $\mathbf{P}$ into the AB space. It follows that $\Delta E^c \geq 0$ if $\bar{A}^c$ is negative semidefinite
which is true if 
$\mathbf{t}$ is negative semidefinite (from Eq.~(\ref{eq:Acdef}), since $\mathbf{B}$ is positive definite).
Following the proof of Ref.~\citenum{scuseria2008ground}, we can diagonalize $\mathbf{t}$ as $\mathbf{t}\mathbf{U}= \mathbf{U} \pmb{\lambda}$.
Note that $\mathbf{t}$ is Hermitian.
The amplitude equation for $\mathbf{t}$ is
\begin{align}
    &\sum_n \frac{1}{n!} \mathbf{B}^{(n)}(\mathbf{t}) = 0 \nonumber \\
    &\mathbf{B} + \mathbf{A}\mathbf{t} + \mathbf{t}\mathbf{A} + 
    \frac{1}{2} [\mathbf{t}\mathbf{B}\mathbf{t} + \mathbf{t}\mathbf{t} \mathbf{B} + \mathbf{B}\mathbf{t}\mathbf{t}] + \ldots = 0
\end{align}
After multiplication from the right and left with $\mathbf{U}_n$ and $\mathbf{U}^\dagger_{n}$ ($n$ denotes the $n$-th eigenvector), this becomes
\begin{align}
    &\mathbf{U}^\dagger_{n} \mathbf{B} \mathbf{U}_n + \mathbf{U}^\dagger_{n}\mathbf{A} \mathbf{t}\mathbf{U}_n + \mathbf{U}^\dagger_{n} \mathbf{t} \mathbf{A}\mathbf{U}_n \nonumber \\
    &+ \frac{1}{2} \mathbf{U}^\dagger_{n} \left[ \mathbf{t}\mathbf{B}\mathbf{t} +  \mathbf{t}\mathbf{t}\mathbf{B} + \mathbf{B}\mathbf{t}\mathbf{t}\right]\mathbf{U}_n+ \dots = 0,
\end{align} 
which is a polynomial of $\lambda_n$
\begin{align}
    &\mathbf{U}^\dagger_{n} \mathbf{B} \mathbf{U}_n + 2 \lambda_n \mathbf{U}^\dagger_{n} \mathbf{A} \mathbf{U}_n \nonumber \\
    &+ \frac{3}{2} \lambda^2_n \mathbf{U}^\dagger_{n} \mathbf{B} \mathbf{U}_n + \dots = 0.
\end{align}
Because $\mathbf{A}$ and $\mathbf{B}$ are positive definite, all the polynomial coefficients are positive. Thus the only possibility for this equation to be satisfied for each real eigenvalue $\lambda_n$ is when $\lambda_n < 0$.
Consequently, we have shown that 
\begin{align}
E^{\mathrm{AB},c}_\mathrm{dRPA} &\geq E^c_\mathrm{dRPA}.
\end{align}
\bibliography{literatur}

\end{document}